\documentclass[twocolumn]{aastex63}
\usepackage{multirow}
\usepackage{amsmath}
\usepackage[figuresright]{rotating}
\usepackage{threeparttable}
\usepackage{subfigure}
\usepackage{epic,eepic}
\usepackage{graphicx}
\usepackage{longtable}
\usepackage{float}
\usepackage{pifont}
\usepackage{hyperref}
\usepackage{gensymb}
\usepackage{textcomp, gensymb}
\usepackage{orcidlink}

\shorttitle{Stellar photospheric UV emission}
\shortauthors{Wang et al.}

\begin{document}

\title{Predicting photospheric UV emission from stellar evolutionary models}

\correspondingauthor{Song Wang}
\email{songw@bao.ac.cn}

\author{Song Wang$^{\orcidlink{0000-0003-3116-5038}}$}
\affiliation{Key Laboratory of Optical Astronomy, National Astronomical Observatories, Chinese Academy of Sciences, Beijing 100101, China}
\affiliation{Institute for Frontiers in Astronomy and Astrophysics, Beijing Normal University, Beijing 102206, China}

\author{Xue Li}
\affiliation{Key Laboratory of Optical Astronomy, National Astronomical Observatories, Chinese Academy of Sciences, Beijing 100101, China}
\affiliation{School of Astronomy and Space Sciences, University of Chinese Academy of Sciences, Beijing 100049, China}

\author{Henggeng Han}
\affiliation{Key Laboratory of Optical Astronomy, National Astronomical Observatories, Chinese Academy of Sciences, Beijing 100101, China}

\author{Jifeng Liu}
\affiliation{Key Laboratory of Optical Astronomy, National Astronomical Observatories, Chinese Academy of Sciences, Beijing 100101, China}
\affiliation{School of Astronomy and Space Sciences, University of Chinese Academy of Sciences, Beijing 100049, China}
\affiliation{Institute for Frontiers in Astronomy and Astrophysics, Beijing Normal University, Beijing 102206, China}
\affiliation{New Cornerstone Science Laboratory, National Astronomical Observatories, Chinese Academy of Sciences, Beijing 100012, People's Republic of China}

\begin{abstract}

Stellar ultraviolet (UV) emission serves as a crucial indicator for estimating magnetic activity and evaluating the habitability of exoplanets orbiting stars.
In this paper, we present a straightforward method to derive stellar photospheric UV emission for F to M main-sequence stars.
By using PARSEC models, we establish relations between near-UV (NUV) and far-UV (FUV) magnitudes from the Galaxy Evolution Explorer (GALEX), NUV magnitudes from the China Space Station Telescope, and stellar effective temperatures and Gaia BP$-$RP color for different metallicities.
Together with the observed sample, we find that for NUV emission, the photospheric contribution to the observed flux is less than 20\% for M stars, around 10\% to 70\% for G stars, and ranges from 30\% to 85\% for G and F stars.
For FUV emission, the photospheric contribution is less than $10^{-6}$ for M stars, below $10^{-4}$ for K stars, around $10^{-4}$ to 10\% for G stars, and between 6\% and 50\% for F stars.
Our work enables the simple and effective determination of stellar excess UV emission and the exploration of magnetic activity.

\end{abstract}



\section{Introduction}

Stellar activity is a valuable tool for probing the strength, distribution, and evolution of magnetic field.
Stars exhibit various activity proxies in different layers of their atmosphere, such as X-ray and radio emission from the corona, ultraviolet (UV) and Ca II H\&K emission from the chromosphere, and spots and flares from the photosphere, etc.
Among these proxies, the UV activity of stars across the HR diagram remains relatively underexplored, possibly due to the significant contamination from the photosphere. 
For late M stars, almost all UV emission can be attributed to stellar activities, while for early M, K, and G stars, the contribution of the photosphere to the near-UV (NUV) emission tends to increase with higher effective temperatures, as the spectral energy distribution shifts toward the blue for hotter stars \citep{2013MNRAS.431.2063S}.
Therefore, previous studies about stellar UV activities are mainly focused on M stars \citep[e.g.,][]{2013MNRAS.431.2063S, 2014AJ....148...64S, 2018AJ....155..122S, 2020ApJ...895....5P}.
It's necessary to provide reasonable estimation of the photospheric UV emission for various types of stars, which can be used to derive excess UV emission from chromosphere by subtracting the photospheric UV flux from the observed flux.
This will allow us to trace the strength and variation of stellar magnetic field and understand the mechanisms of stellar magnetic dynamo.

On the other hand, stellar UV emission can critically affect the chemistry in
the planet atmospheres in a way which is crucial for life.
For example, the extreme-UV (100--900 \AA) and X-ray radiation (2--100 \AA) can heat and evaporate the primary hydrogen-rich planetary atmosphere \citep{2003ApJ...598L.121L,2009A&A...496..863C} and influence the production of $\rm{O_{3}}$ in the planet atmosphere \citep{2003AsBio...3..689S}, while the NUV radiation  ($\sim$2000--3000 \AA) can trigger the formation of organic molecules \citep[e.g.,][]{2009Natur.459..239P}. 
A UV habitable zone has been defined as the region where a planet receives moderate UV radiation \citep{2007Icar..192..582B, 2023MNRAS.522.1411S} conducive for biochemical processes and simultaneously preventing damage to biological systems \citep{2007Icar..192..582B}. 
More studies on stellar UV emission can enhance our understanding of the mass loss processes of planets \citep{2007A&A...461.1185L,2008A&A...479..579P}, the chemistry of planetary atmospheres, and their suitability for generating and sustaining life \citep{2004Icar..171..229S}.

Aiming to simply derive the excess UV emission from chromosphere and probe stellar UV activity for main-sequence stars, in this paper, we try to provide reasonable estimates of photospheric UV emission for different types of stars based on stellar models and observations.
The paper is organized as follows.
In Section \ref{data.sec}, we introduce the dataset (i.e., the observational UV sample of different types of stars) and the method to derive the photospheric emission.
Section \ref{result.sec} presents the results, including the relations between far-UV (FUV) and NUV magnitudes and stellar effective temperatures and color across different metallicities.
In addition, we discuss the ratio of photospheric to observed UV emission, and the relation between the NUV magnitude from China Space Station Telescope (CSST) and stellar temperature and color.
Finally, a short summary is given in Section \ref{sum.sec}.

\section{Dataset and Method}
\label{data.sec}

The Galaxy Evolution Explorer (hereafter GALEX) has conducted an all-sky survey in the FUV ($\lambda_{\rm eff}\sim1528$ \AA; 1344--1786 \AA) and NUV ($\lambda_{\rm eff}\sim2310$ \AA; 1771--2831 \AA) bands \citep{2007ApJS..173..682M}, making it an invaluable resource for our study.
In this work, we used the GR6$+$7 catalog \citep{2017ApJS..230...24B}, which includes magnitudes for a total of 82,992,086 objects derived from an All-Sky Imaging Survey and a Medium-depth Imaging Survey.

Recently, \citet{2024ApJ...966...69L} carried out a statistical analysis of stellar activity of about 5900 M stars using the GALEX catalog and the Large Sky Area Multi-Object Fiber Spectroscopic Telescope (LAMOST) data, while X. Li et al. (2024, in prep.) performed a further analysis of about 1.1 millions of F, G, and K stars  with stellar parameter estimations.
These authors cross-matched the GALEX {\it PhotoObjAll} catalog and the LAMOST DR9 low-resolution catalog to derive a comprehensive sample of objects with UV emission.
They ran a series of steps to classify their sample into various categories, including binaries, young stellar objects, dwarfs and giants, and non-stellar sources \citep[See][for more details]{2024ApJ...966...69L}.  
Then, using the distance measurements from Gaia EDR3 \citep{2021AJ....161..147B} and extinction data from the Pan-STARRS1 3D dust map \citep{2015ApJ...810...25G}, they calculated NUV and FUV luminosities for single dwarf stars and giants.
Using their data sets, we decided to studied the photospheric UV emission for F, G, K and M dwarfs.

\begin{figure}[ht!]
\begin{center}
\includegraphics[width=0.48\textwidth]{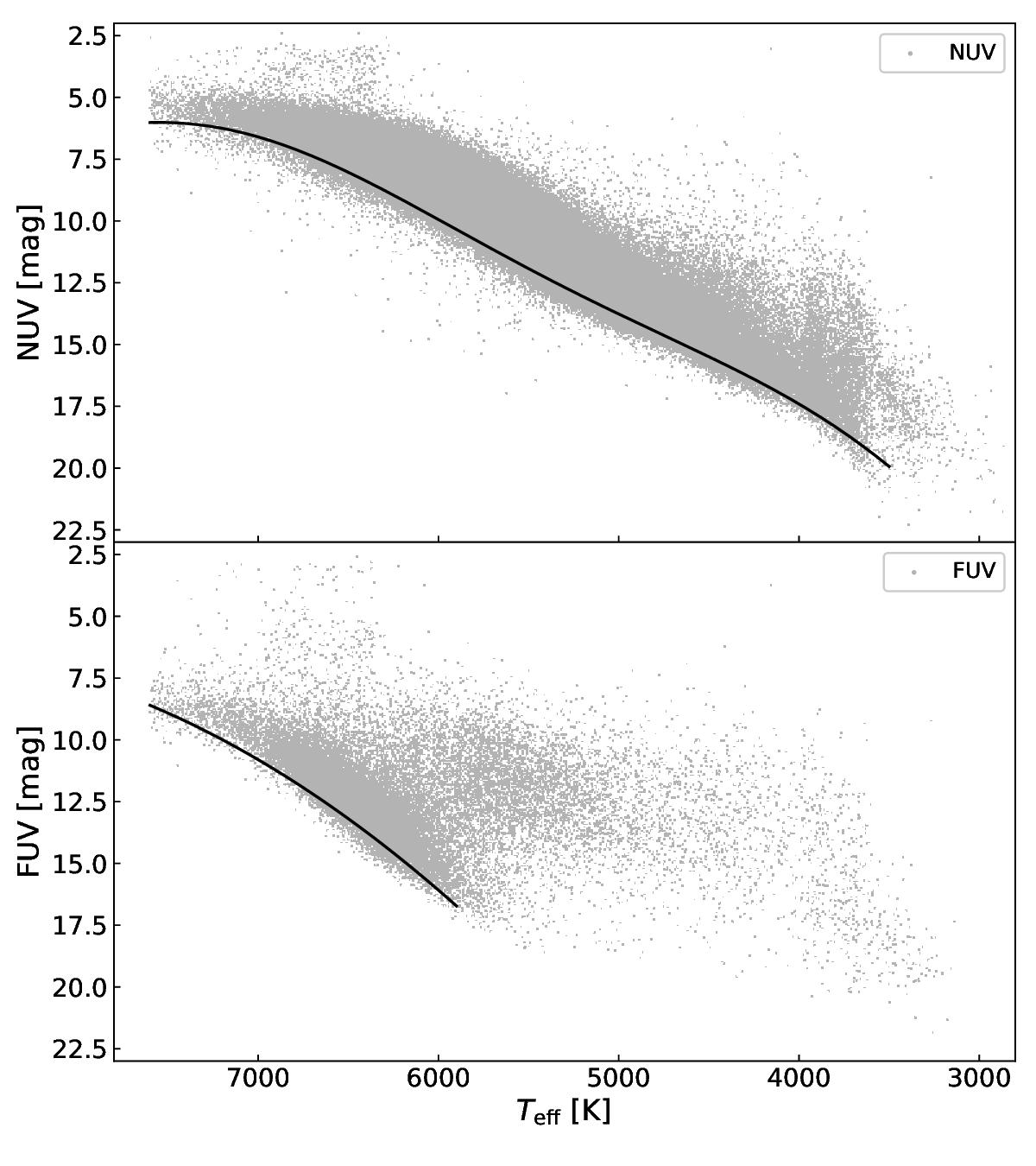}
\caption{Observed NUV and FUV magnitudes for our sample. The black lines are the fitted baselines using most inactive stars (i.e., with faintest UV emission).
\label{base.fig}}
\end{center}
\end{figure}

\begin{figure*}
\begin{center}
\includegraphics[width=0.48\textwidth]{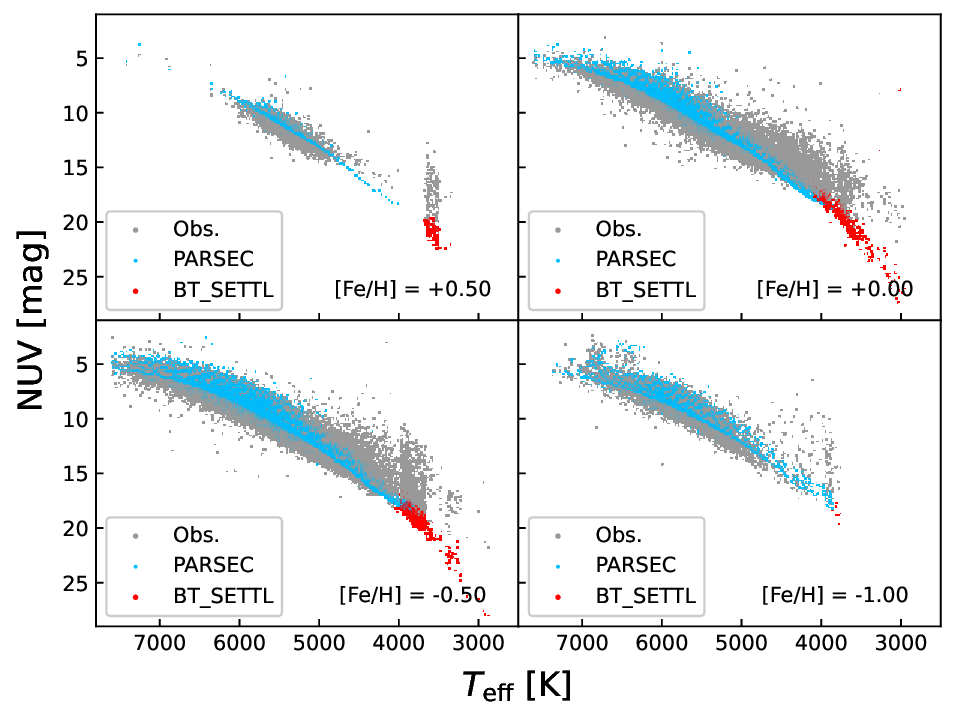}
\includegraphics[width=0.48\textwidth]{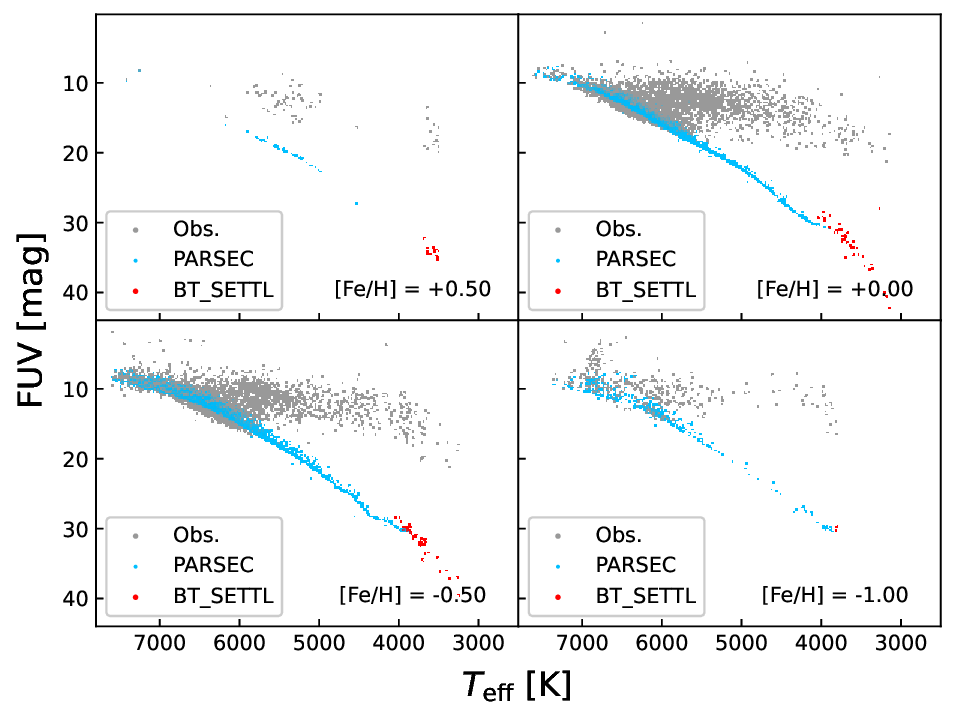}
\includegraphics[width=0.48\textwidth]{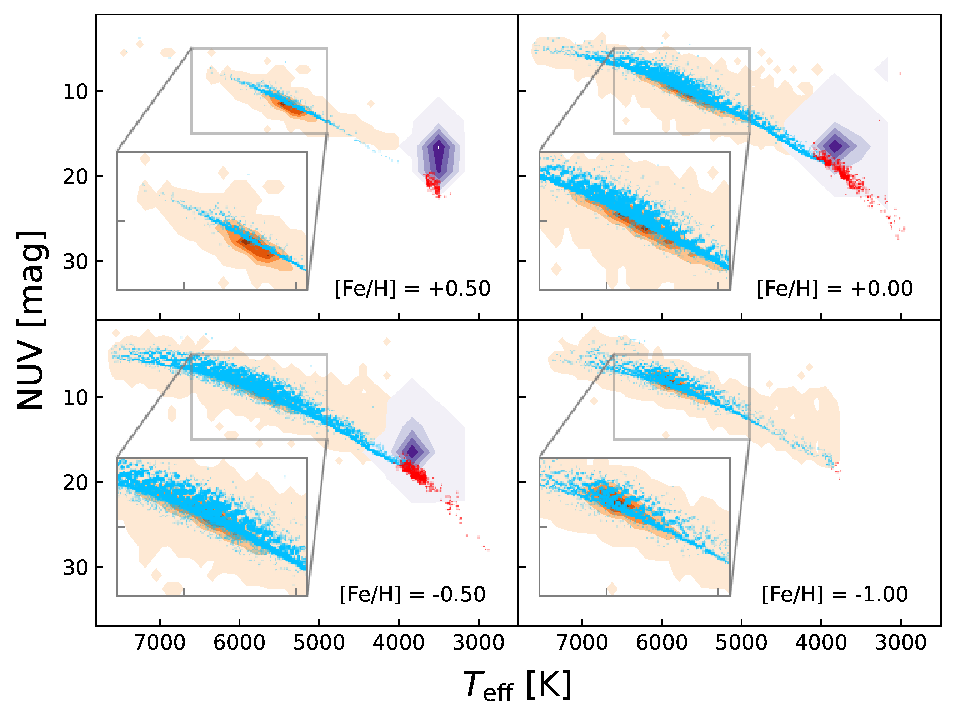}
\includegraphics[width=0.48\textwidth]{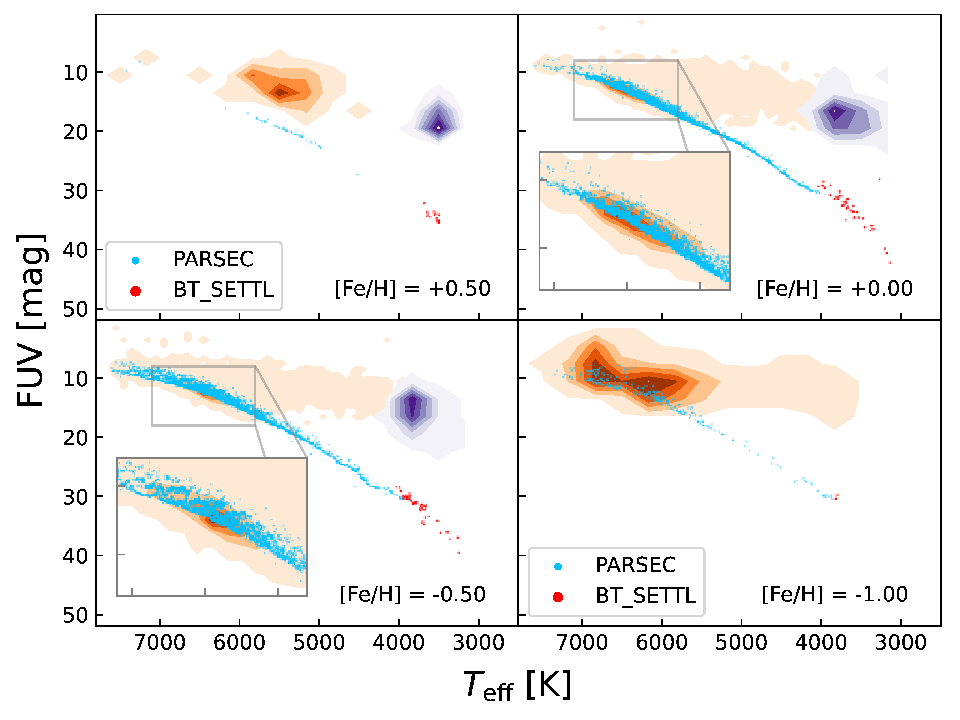}
\caption{Left panels: Observed NUV magnitudes (grey points) compared with photospheric NUV magnitudes derived from PARSEC models (blue points; F to K stars) and BT-SETTL models (red points; M stars) using atmospheric parameters (i.e., $T_{\rm eff}$, log$L$, [Fe/H]). The observed magnitudes are shown with  grey points in the top panel and contours (i.e., orange contours for F to K stars while violet contours for M stars) in the bottom panel, respectively. 
Right panels: Observed FUV magnitudes (grey points) compared with photospheric FUV magnitudes derived from PARSEC models (blue points; F to K stars) and BT-SETTL models (red points; M stars).
\label{tefflogg.fig}}
\end{center}
\end{figure*}

Generally, there are two methods to derive the photospheric contribution to the UV band:
  \begin{itemize}
\item Using the observed sample to establish a baseline for UV emission. 
Figure \ref{base.fig} shows the FUV and NUV magnitudes for different types of stars. The black lines represent the fitting of the baselines using most inactive stars.
It can be seen that due to the faintness of UV emissions from cooler stars, such as K and M stars, only the UV-luminous sources are observable. 
For example, the FUV sample for stars cooler than 6000 K is seriously incomplete, while the NUV sample for stars cooler than 3500 K is incomplete as well, suggesting that constructing reliable baseline values for these stars is difficult.
Therefore, it's unreliable to derive photospheric UV emission using the observed sample of late-type stars.
\item Using atmospheric parameters to search for the best-fit stellar model to determine the photospheric UV emission. Generally, the surface temperature, surface gravity (or luminosity), and metallicity are used to find the best model in a grid of isochrones \citep[e.g.,][]{2013MNRAS.431.2063S,2020ApJ...902..114W,2024ApJ...966...69L}.
On the one hand, searching through models for millions of stars are time-consuming.
On the other hand, due to the uncertainties in parameter measurements and the sparse grids of stellar models, in many cases a best-fit model is unavailable by matching the observed atmospheric parameters with model grids.
Figure \ref{tefflogg.fig} shows that the photospheric emissions from model fitting do not align along a tight baseline, possibly due to uncertainties in parameter measurements.

  \end{itemize}

In order to devise a straightforward method to determine stellar photospheric UV emission from models, we collected a group of popular stellar evolutionary models, including 
BaSTI \citep[a Bag of Stellar Tracks and Isochrones;][]{2018ApJ...856..125H}\footnote{http://basti-iac.oa-abruzzo.inaf.it/isocs.html}, PARSEC \citep[PAdova and TRieste Stellar Evolution Code;][]{2012MNRAS.427..127B}\footnote{http://stev.oapd.inaf.it/cgi-bin/cmd}, 
MIST \citep[MESA Isochrones and Stellar Tracks;][]{2016ApJ...823..102C}\footnote{https://waps.cfa.harvard.edu/MIST/model\_grids.html}, 
and BT-SETTL \citep{2011ASPC..448...91A}\footnote{http://svo2.cab.inta-csic.es/theory/iso3}.
The UV magnitudes of the main-sequence stars in these models will be used to trace stellar photospheric emission.
We downloaded these models with metallicities of [Fe/H] $=$ 0.5, 0, $-$0.5, and $-$1.0. Note that although the PARSEC models use [M/H] as the measure of metallicity, we treated [M/H] as [Fe/H], ignoring any enhancement of $\alpha$ elements.

\section{Results and Discussion}
\label{result.sec}

\begin{figure*}[t]
\begin{center}
\includegraphics[width=0.48\textwidth]{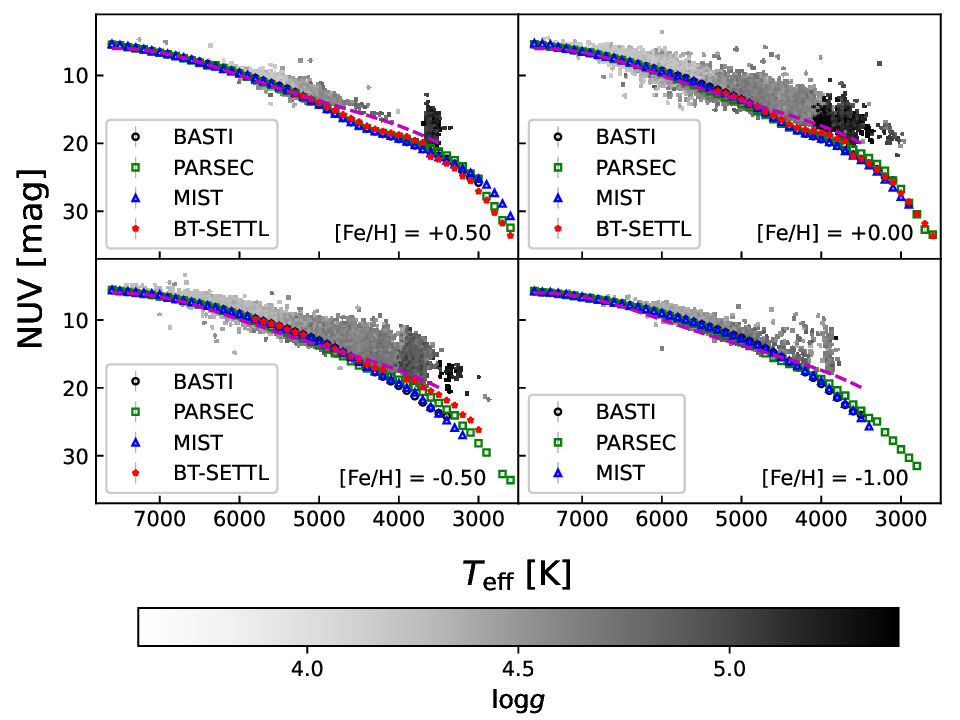}
\includegraphics[width=0.48\textwidth]{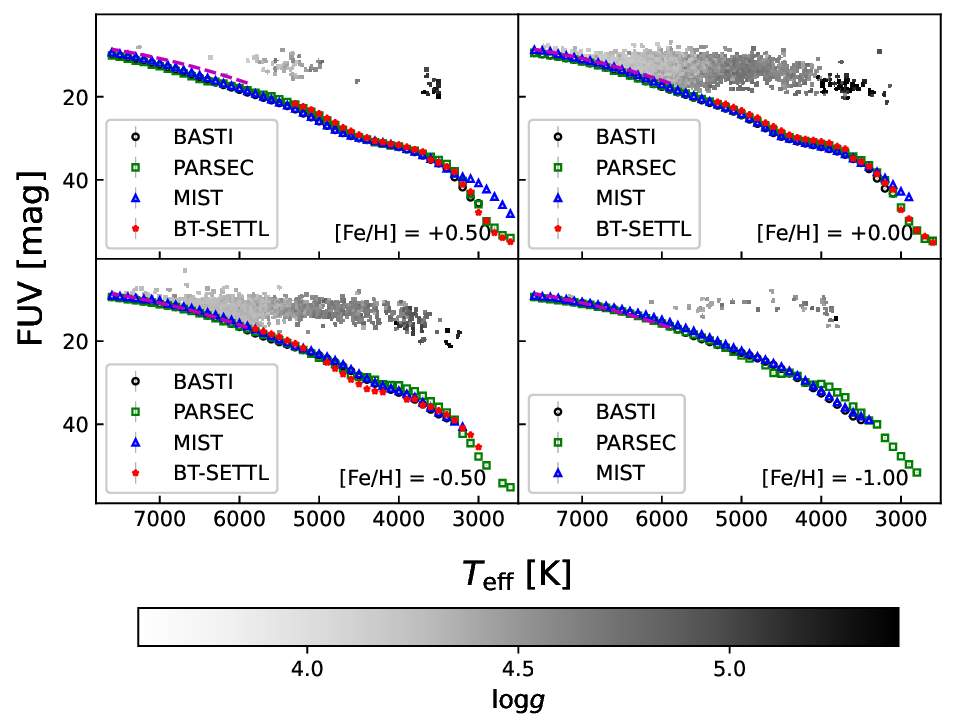}
\includegraphics[width=0.48\textwidth]{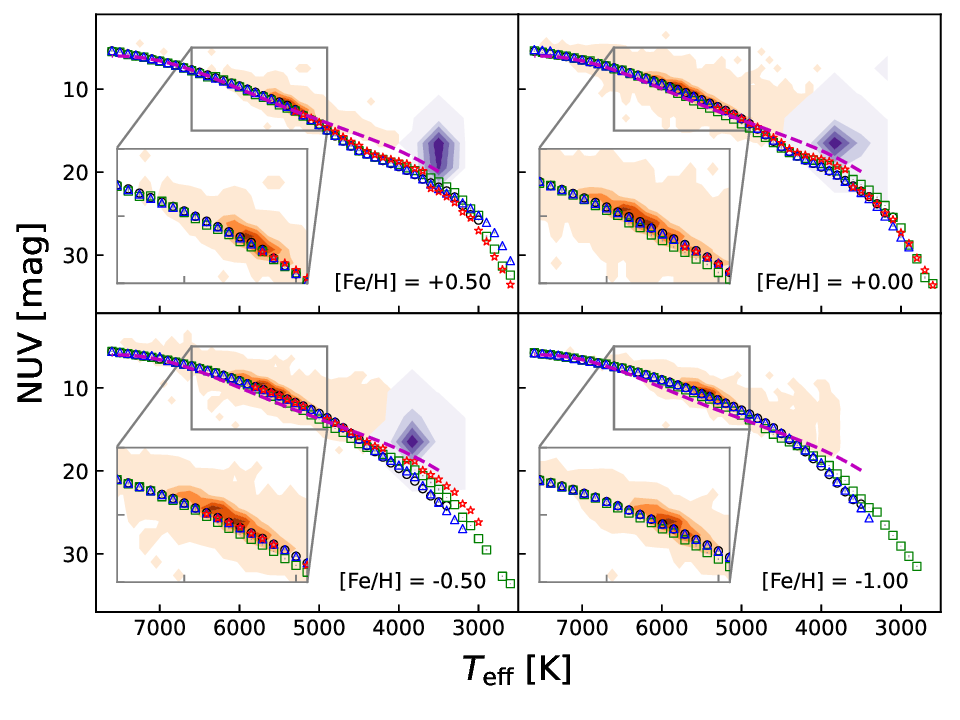}
\includegraphics[width=0.48\textwidth]{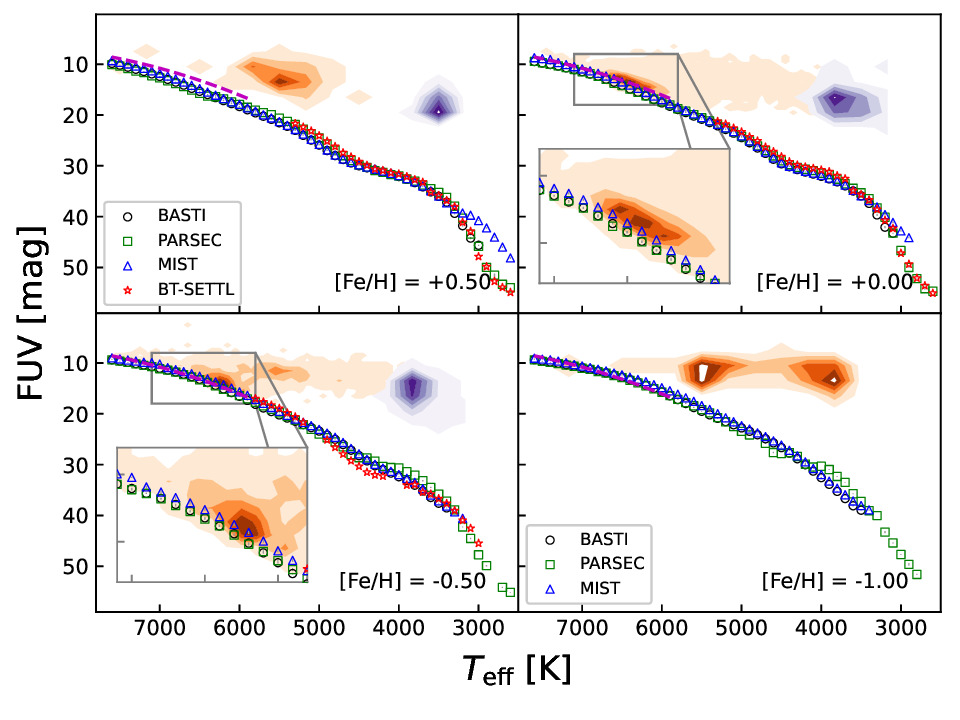}
\caption{Left panels: Observed NUV magnitudes (points) compared with photospheric NUV magnitudes predicted by different models (e.g., BASTI, PARSEC, MIST, BT-SETTL). 
The colors of points in the top panel indicate temperatures. The observed magnitudes are shown with points in the top panel and contours (i.e., orange contours for F to K stars while violet contours for M stars) in the bottom panel, respectively. 
The model magnitudes are calculated as the median magnitudes within a temperature bin of 200 K.
The magenta dashed lines are the fitted baselines using the observed sample in Figure \ref{base.fig}.
Right panels: Observed FUV magnitudes (points) compared with photospheric FUV magnitudes predicted by different models (e.g., BASTI, PARSEC, MIST, BT-SETTL). 
\label{model.fig}}
\end{center}
\end{figure*}

Figure \ref{model.fig} shows a comparison of UV magnitudes between observation and different stellar models, including BASTI, PARSEC, MIST, and BT-SETTL.
It's evident that the models closely match the lower bounds of the observed magnitudes. 
The corner plots show that the majority of the stars exhibit UV emission consistent with model predictions, suggesting that a considerable portion of UV emission of these stars, especially G and F stars, originates from their photospheres.
On the one hand, the BASTI, PARSEC, and MIST models are consistent for stars with temperature higher than 4000 K (i.e., F to K stars).
On the other hand, for cooler stars, the BT-SETTL models are in better agreement with the PARSEC models than the BASTI and MIST models.
In general, the BT-Settl model, which uses revised solar abundances and updated atomic and molecular line opacities, can well reproduce the observed spectra of cool dwarfs \citep{2013A&A...556A..15R} and is considered appropriate for describing the emission from cool stars.
Consequently, the PARSEC models are considered suitable for F to M stars, and they will be used to establish a baseline of UV magnitudes representing stellar photospheric UV emission.

\begin{figure*}
\begin{center}
\includegraphics[width=0.32\textwidth]{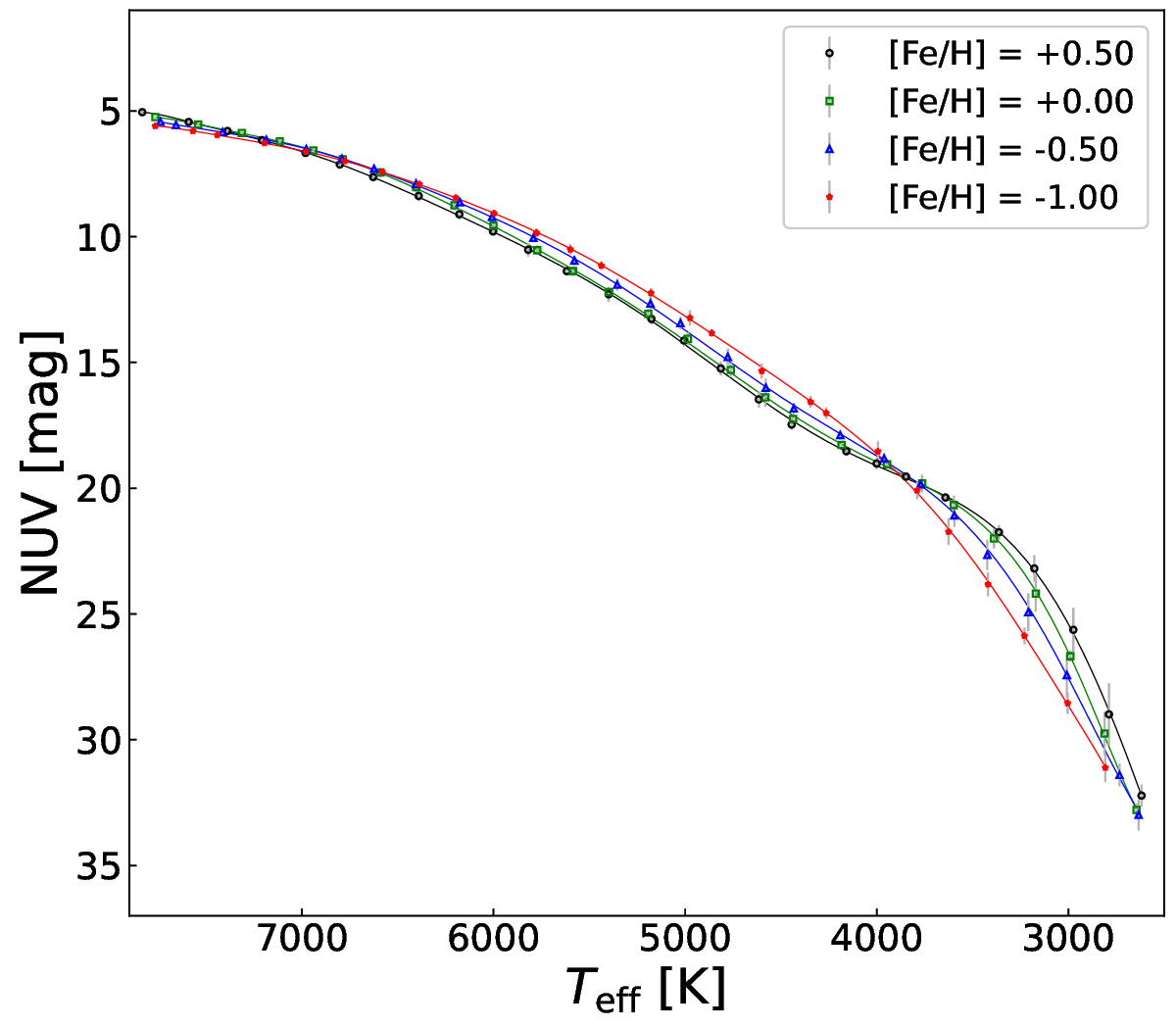}
\includegraphics[width=0.32\textwidth]{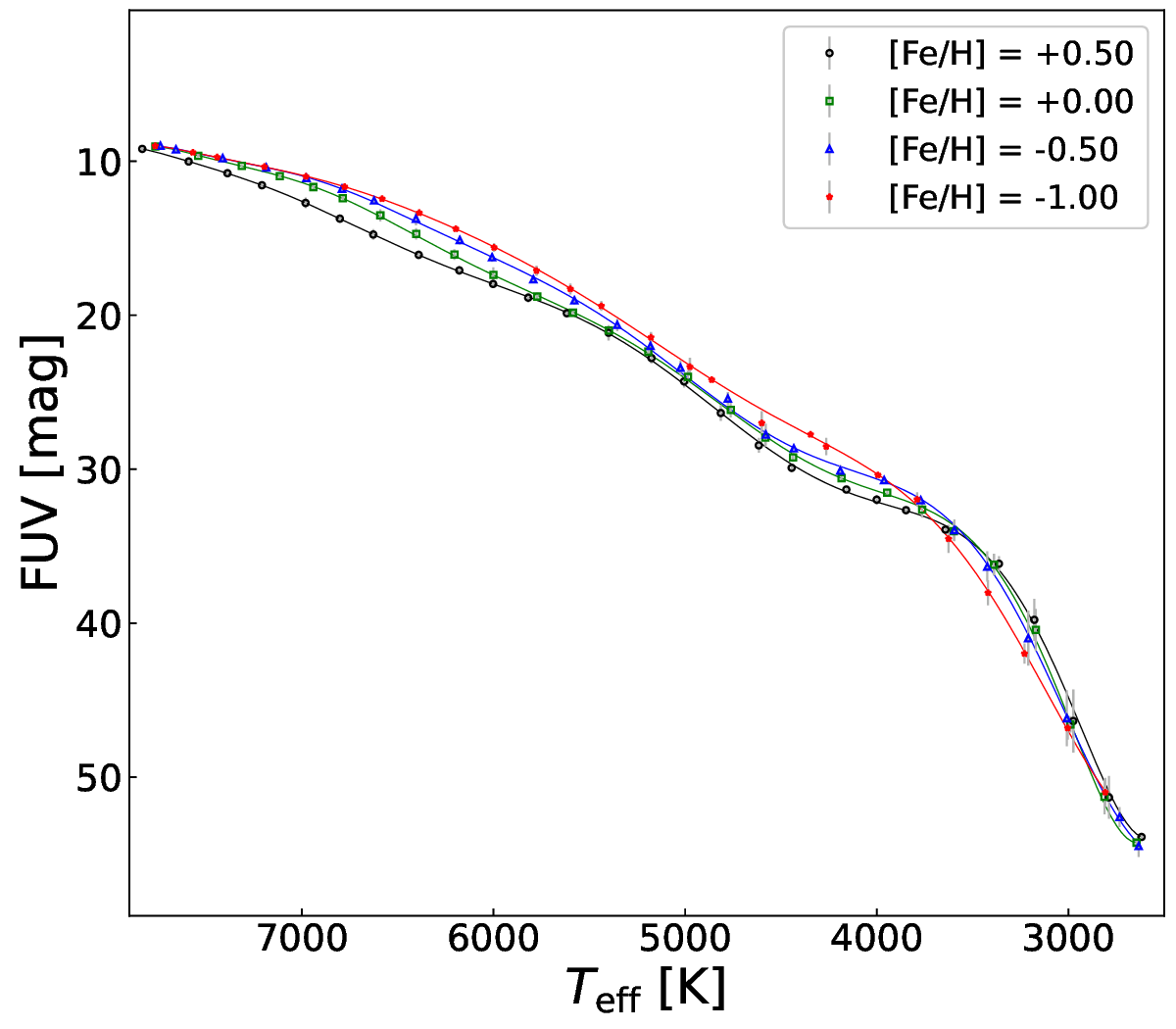}
\includegraphics[width=0.32\textwidth]{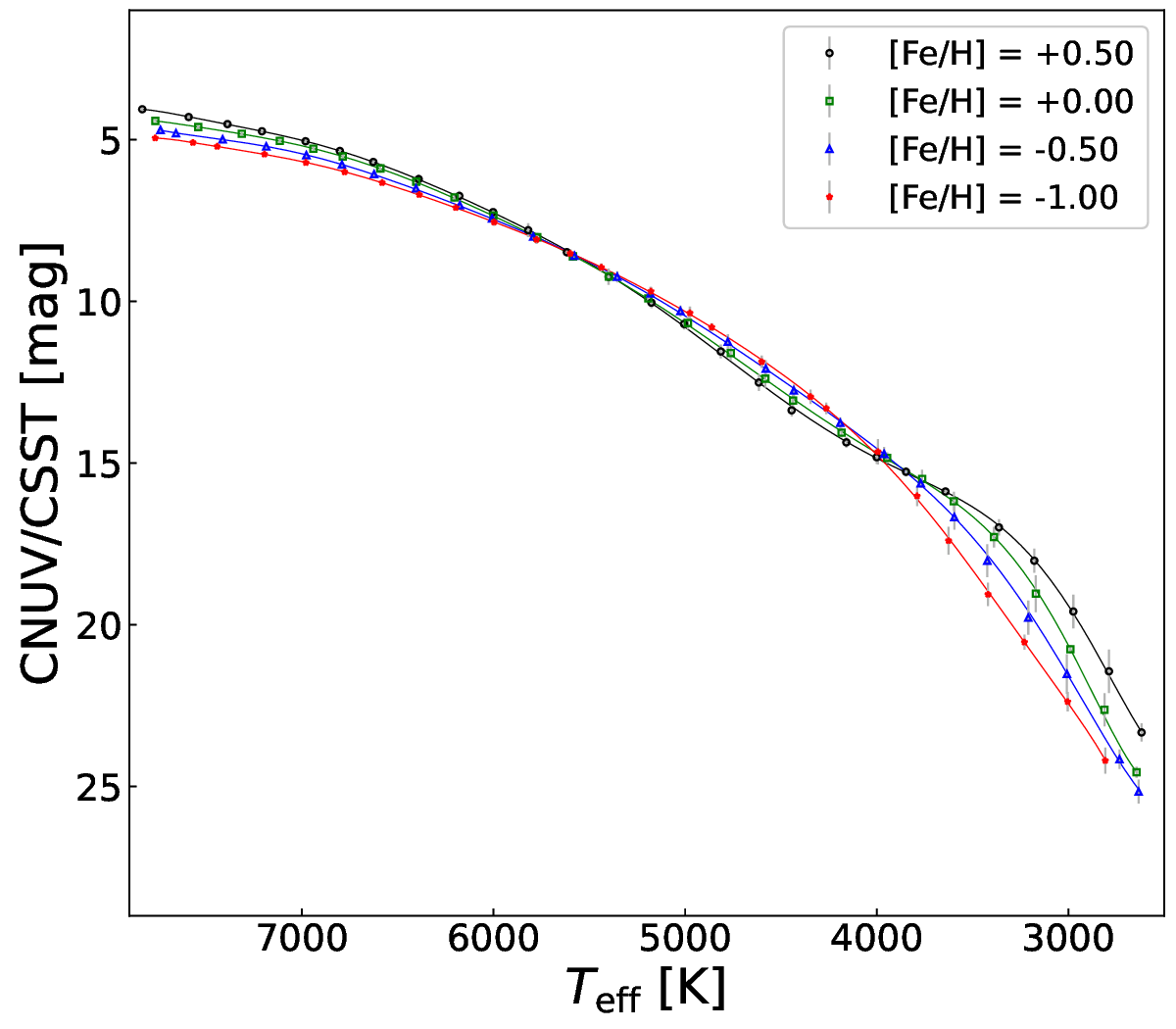}
\includegraphics[width=0.32\textwidth]{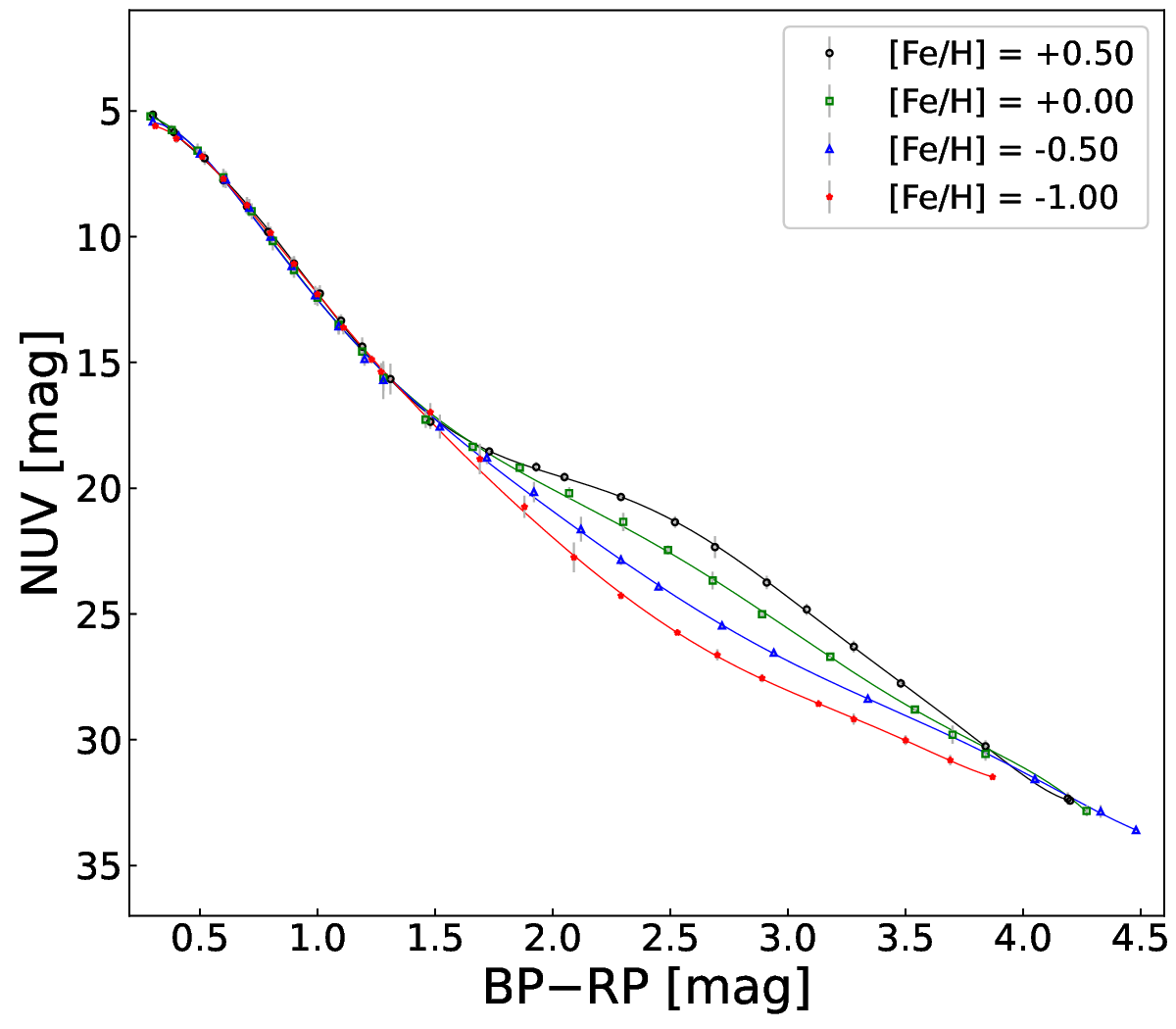}
\includegraphics[width=0.32\textwidth]{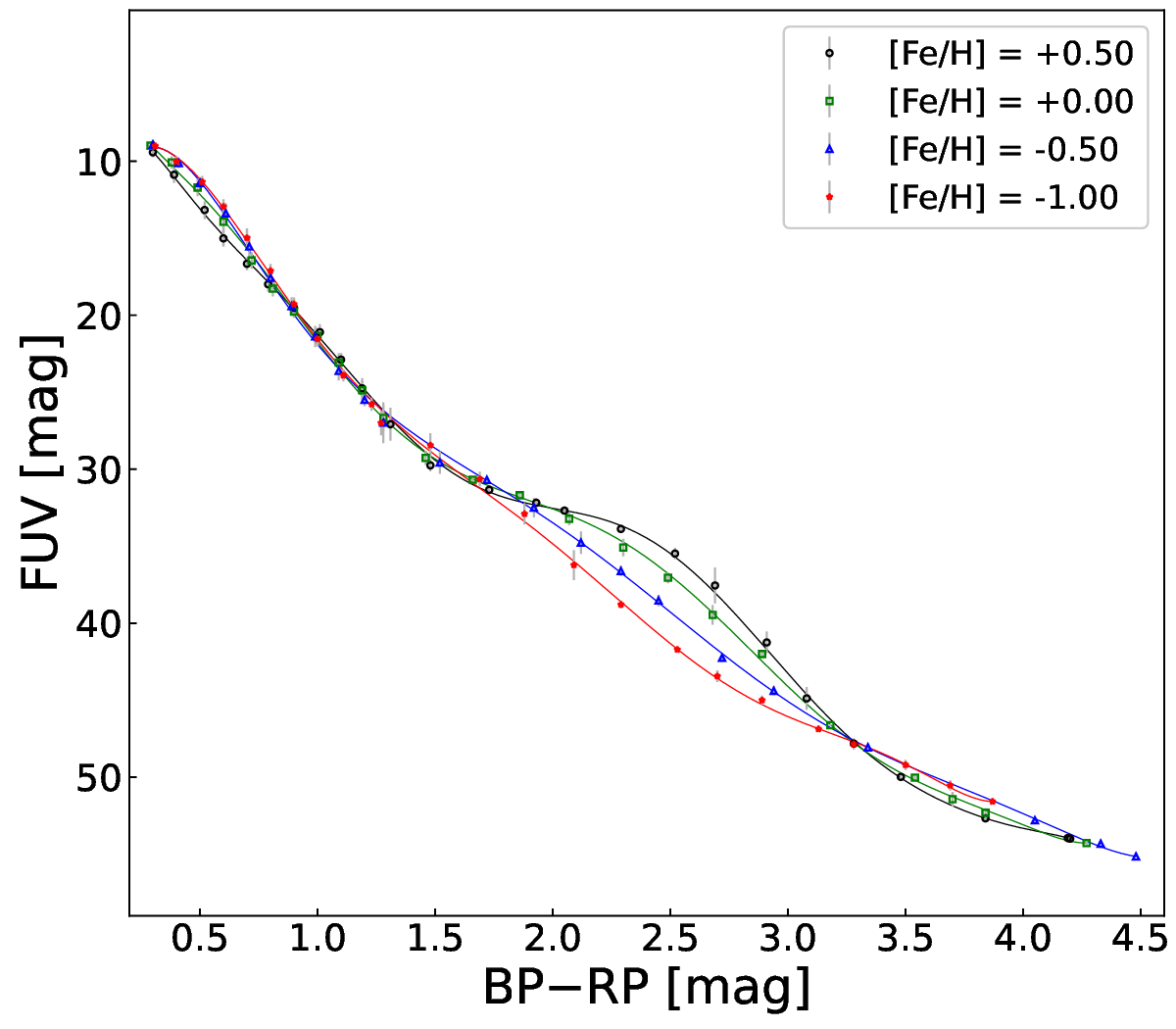}
\includegraphics[width=0.32\textwidth]{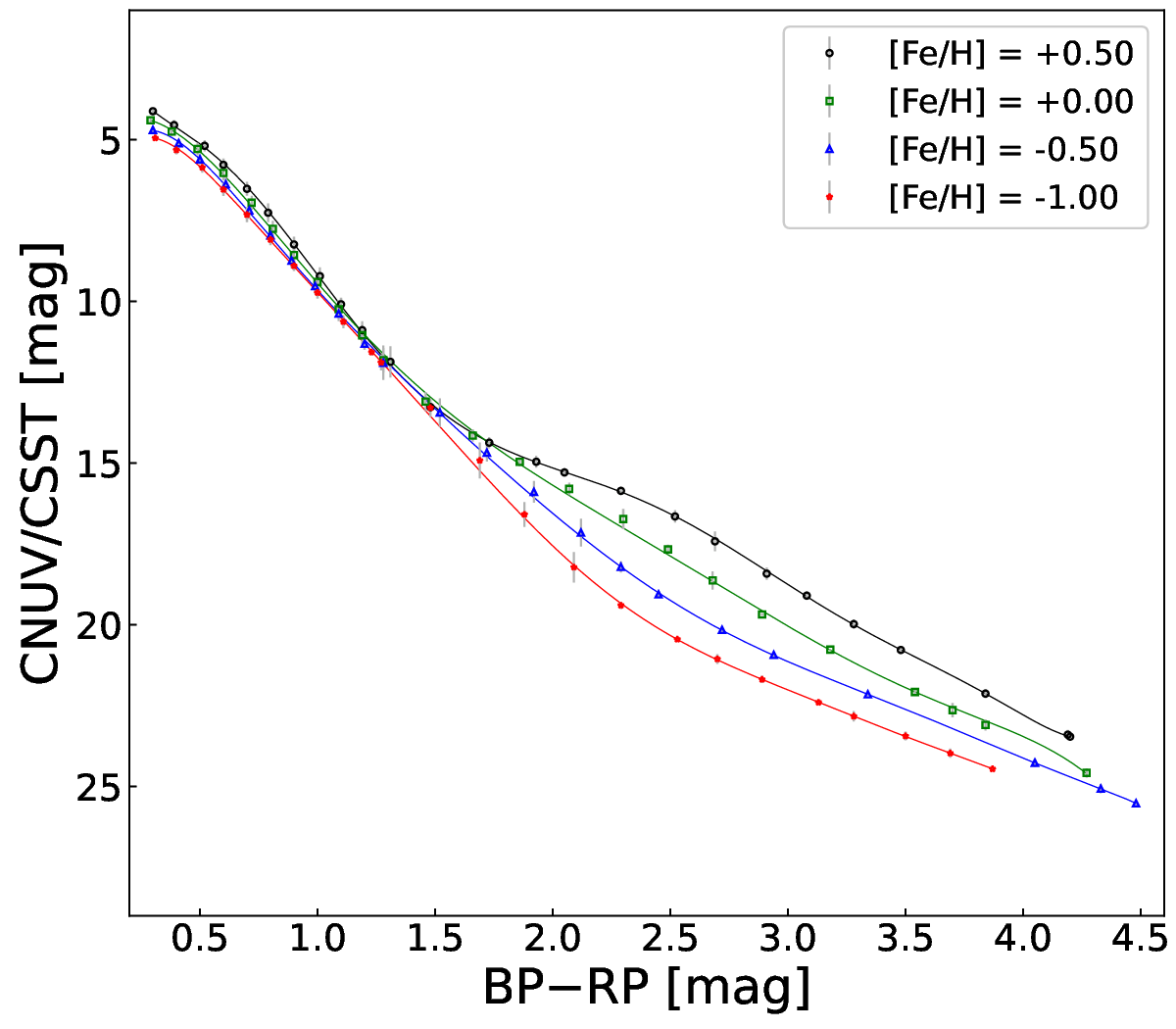}
\caption{Top panels: Fitting results of the relations between NUV/GALEX (left panel), FUV/GALEX (middle panel), and NUV/CSST (right panel) magnitudes and stellar effective temperature $T_{\rm eff}$. 
Bottom panels: Fitting results of the relations between NUV/GALEX (left panel), FUV/GALEX (middle panel), and NUV/CSST (right panel) magnitudes and stellar color BP$-$RP. 
\label{fit.fig}}
\end{center}
\end{figure*}

We employed a tenth-order polynomial to model the relationship between UV magnitude and effective temperature for dwarfs with different [Fe/H] values:
\begin{equation}
    m_{\rm UV} = \sum_{i=0}^{i=10} a_i (\frac{T_{\rm eff}}{1000\ K})^{10-i}.
\label{uv_teff.eq}
\end{equation}
The fitting coefficients $a_i$ ($i = $1, 2, $\cdots$, 10) for NUV and FUV bands are provided in Table \ref{nuvfit.tab}. The data points used for fitting in Figure \ref{fit.fig} (Top panels) are the averaged magnitudes of models within a temperature bin of 200 K.

Simultaneously, photometric surveys offer a substantially larger number of objects compared to spectroscopic surveys. 
For example, Gaia DR3 provided $G$-band photometry for 1.8 billion sources and BP- and RP-band photometry for 1.5 billion sources \citep{2023A&A...674A...1G}, whereas LAMOST DR11, the largest stellar spectra database until now, only provided spectra for about 10 million stars.
Therefore, it is necessary to establish relations between photospheric UV emissions and stellar colors.
We also employed polynomial fittings to derive relations between photospheric UV magnitude and BP$-$RP color for F to M dwarfs, using different orders for different metallicities.
The BP$-$RP color was also derived from PARSEC model.
The fitting equation is as follows,
\begin{equation}
    m_{\rm UV} = \sum_{i=0}^{i=10} a_i ({\rm BP-RP})^{10-i}.
\label{uv_teff.eq}
\end{equation}
The data points in Figure \ref{fit.fig} (Bottom panels) are the averaged magnitudes of models within a color bin of 0.1 mag for BP$-$RP $\leq$ 1.2 mag and 0.2 mag for BP$-$RP $>$ 1.2 mag.
The fitting coefficients are listed in Table \ref{nuvfitc.tab}.

\subsection{Ratio of photospheric to observed UV emission}

\begin{figure*}
\begin{center}
\includegraphics[width=0.48\textwidth]{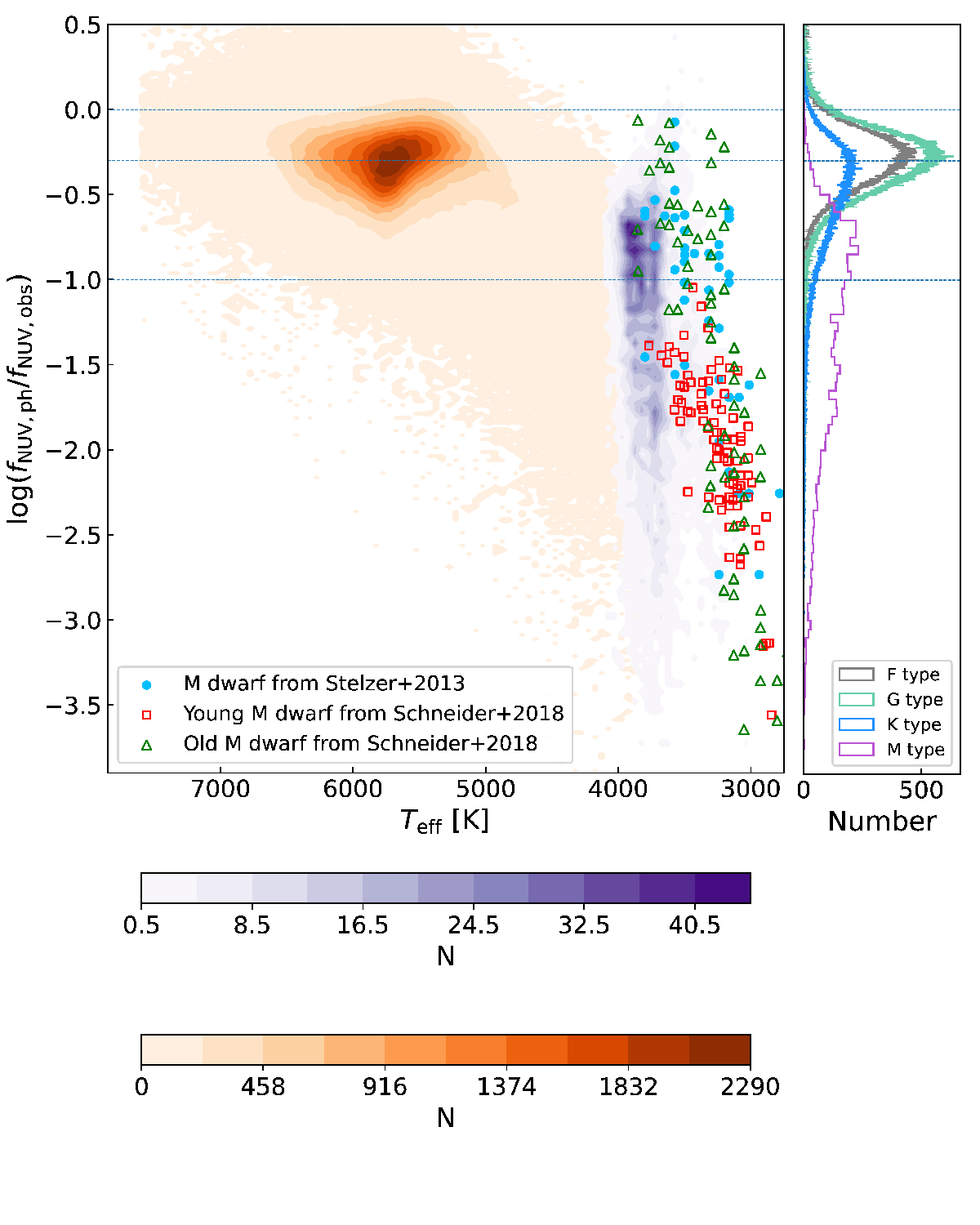}
\includegraphics[width=0.48\textwidth]{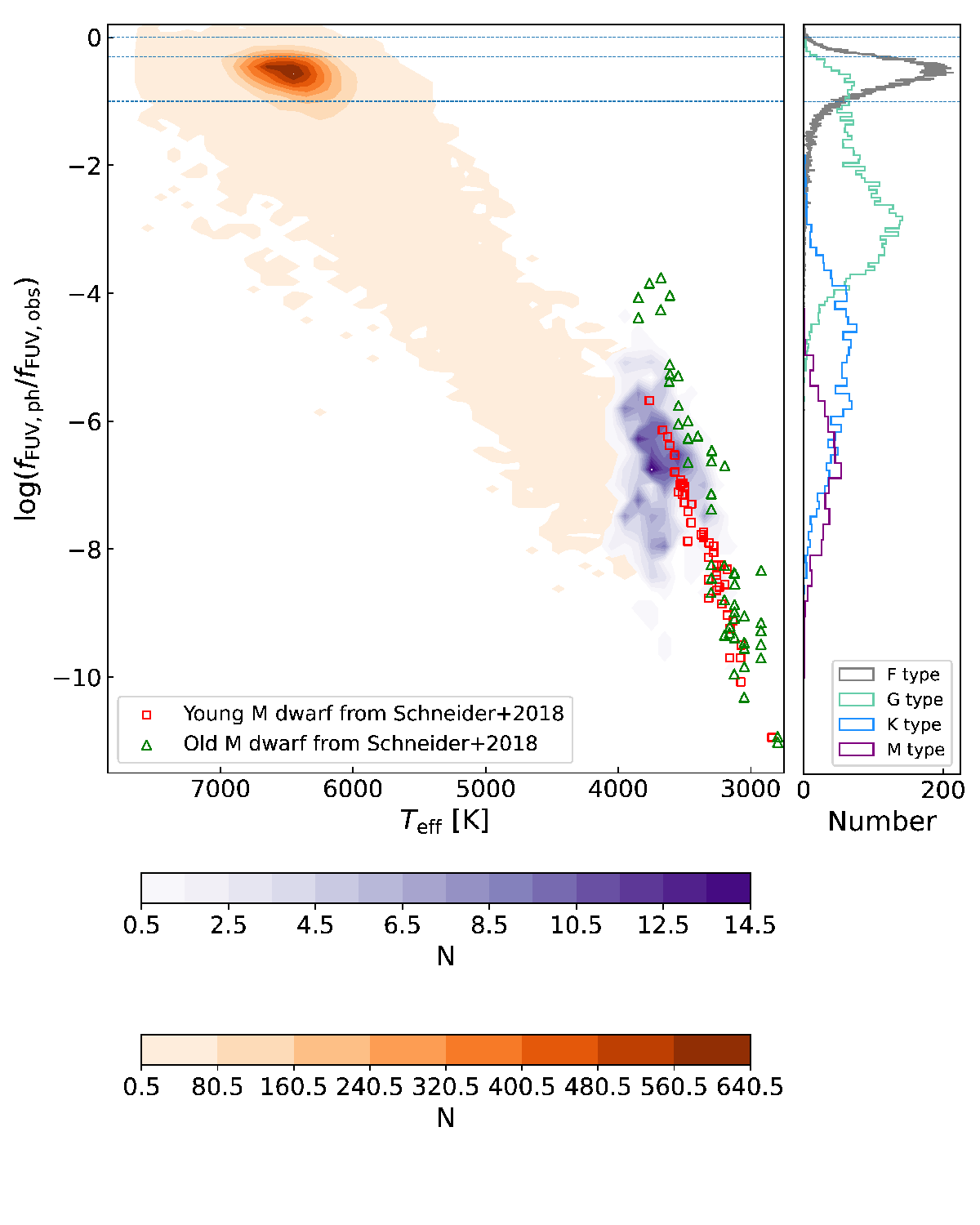}
\caption{Left panel: Ratio of photospheric NUV emission to total NUV emission. 
The orange contours represent the number distribution of F to K stars, while the violet contours represent the number distribution of M stars. The blue points and their data are from \citet{2013MNRAS.431.2063S}. The red squares and green triangles represent young and old M dwarfs from \cite{2018AJ....155..122S}, respectively.
The dotted horizontal lines (from top to bottom) mark ratios of 1, 0.5, and 0.1, meaning 100\%, 50\%, and 10\% of the UV emission is from the photosphere, respectively.
The colorbars indicate the number of stars.
Histograms are included as the subplot to show the distributions of the ratios for F, G, K, and M stars, with sample sizes of 183867, 499829, 112779, 5797, respectively. For clarity, the bin sizes differ across stellar types: 0.001 for F stars, 0.0005 for G stars, 0.00125 for K stars, and 0.05 for M stars.
Right panel: Ratio of FUV photospheric emission to total emission. Histograms are included as the subplot to show the distributions of the ratios for F, G, K, and M stars, with sample sizes of 21603, 5397, 1157, 448, respectively. For clarity, the bin sizes differ across stellar types: 0.006 for F stars, 0.06 for G stars, 0.12 for K stars, and 0.24 for M stars.
\label{ratio.fig}}
\end{center}
\end{figure*}

By subtracting the photospheric UV flux from the observed flux, we can derive the excess UV flux emitted by the chromosphere, which is related to stellar magnetic activity.
Figure \ref{ratio.fig} shows the fraction of photospheric emission in the total emission for the NUV/GALEX and FUV/GALEX bands.
A ratio of zero means the observed UV emission is entirely from the chromosphere, while a ratio of one means all UV emission is from the photosphere.

In the NUV band, for most M stars, the photosphere contributes less than $\approx$20\% of the total NUV emission (Table \ref{uvph.tab}), suggesting the dominant contribution to the NUV emission comes from the chromosphere, as reported by \citet{2013MNRAS.431.2063S}. 
Recently, \cite{2018AJ....155..122S} measured photospheric contribution for a sample of young ($<$50 Myr) and old ($\approx$5 Gyr) M stars.
They found that for the old sample, the average photospheric contribution is about 26\% when $T_{\rm eff} >$ 3200 K, and around 0.7\% for cooler stars. For the young sample, the average photospheric contribution is approximately 2\% when $T_{\rm eff} >$ 3200 K, and similarly about 0.7\% for cooler stars.
Our M-dwarf sample mainly consists of field dwarfs with temperatures above 3200 K.
The distributions of photospheric contribution of (old) M dwarfs in  \citet{2013MNRAS.431.2063S} and \cite{2018AJ....155..122S} are in good agreement with our results (Figure \ref{ratio.fig}).

On the other hand, the higher the effective temperature, the larger the average contribution of the photosphere to the NUV emission. 
For K stars, the photospheric contribution to NUV emission typically ranges from 10\% to 70\% (10th to 90th percentiles), while for G and F stars, this range is approximately 30\% to 85\% (Table \ref{uvph.tab}).
The median photospheic NUV contribution is about 6\%, 30\%, 50\%, and 50\% for M, K, G, and F stars, respectively.

In the FUV band, the photospheric contribution for nearly all M stars is less than $2 \times 10^{-6}$, indicating almost all of the emission comes from the chromosphere.
\cite{2018AJ....155..122S} reported a range of $10^{-11}$ to $10^{-3}$ for the photospheric contribution in M stars.
For K stars, the photospheric FUV contribution remains low, ranging from $10^{-7}$ to $2 \times 10^{-4}$ (10th to 90th percentiles).
For G stars, the contribution ranges from $10^{-4}$ to 10\%, while for F stars, it spans from 6\% to 50\%, indicating a significant contribution from the photosphere.
The median photospheic FUV contribution is about $2\times10^{-7}$, $7\times10^{-6}$, $2\times10^{-3}$, and 25\% for M, K, G, and F stars, respectively.

\begin{figure}
\begin{center}
\includegraphics[width=0.48\textwidth]{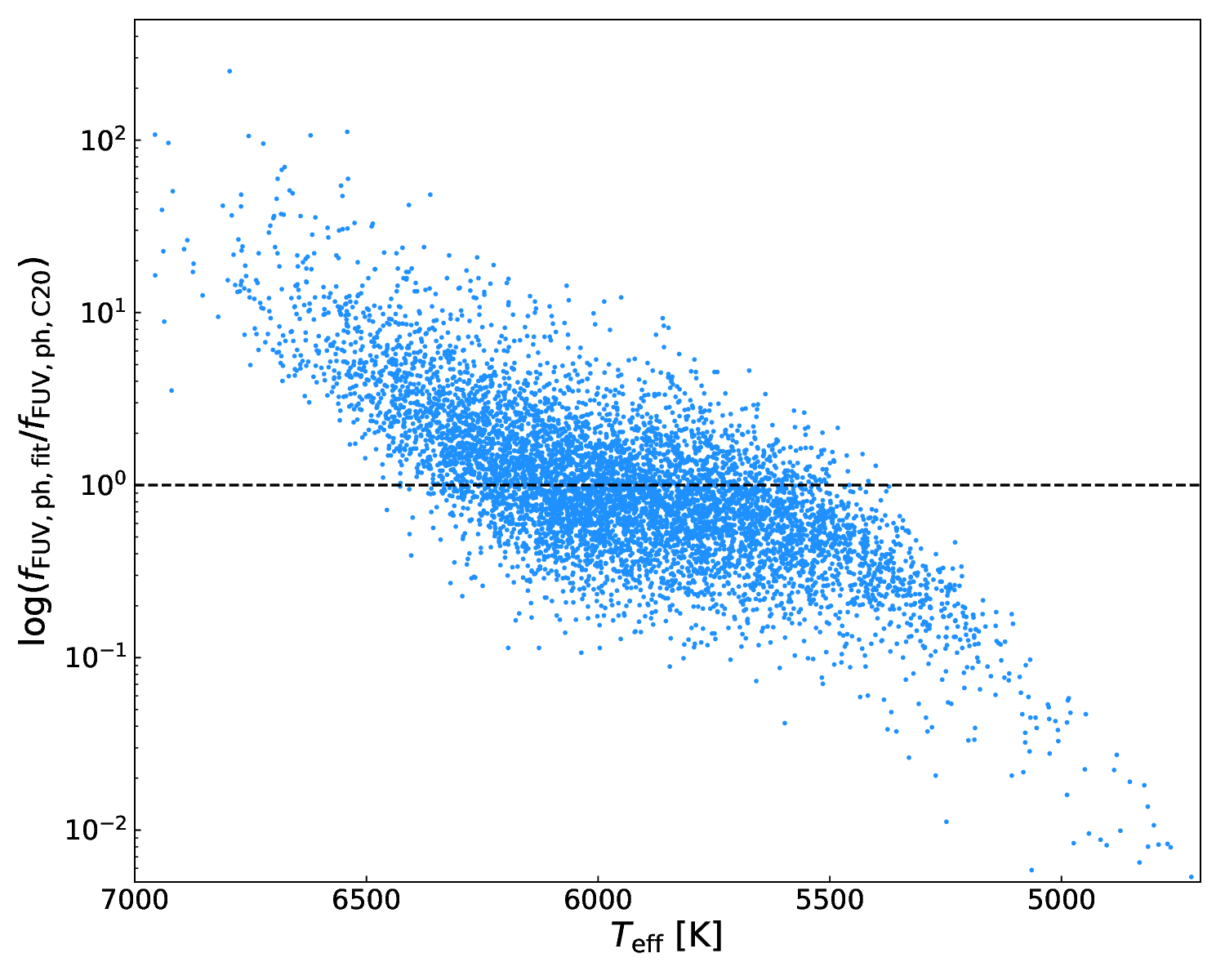}
\caption{Comparison of photospheric FUV emission estimated using the method presented in this paper and that from \cite{2020AJ....160..217C}.
\label{c20.fig}}
\end{center}
\end{figure}

Using a sample of dwarfs, \cite{2020AJ....160..217C} provided an equation to estimate the photospheric FUV contribution: ${\rm FUV} - B = -28.70(B-V)^2 + 48.16(B-V) - 6.35$ for stars with $B-V$ color between 0.55 and 0.8.
We cross-matched our sample with the fourth United States Naval Observatory CCD Astrograph Catalog to obtain $B$ and $V$ magnitudes and calculated the photospheric FUV magnitudes following their method.
Figure \ref{c20.fig} shows that for stars with 5500 K $\lesssim T_{\rm eff} \lesssim$ 6500 K, the two results are consistent. 
However, for hotter and cooler stars, noticeable discrepancies arise, likely due to the inappropriate application of their equation in these temperature ranges, as their sample may be incomplete in these regions (similar to Figure \ref{base.fig}).

The scatter of the distribution (Figure \ref{ratio.fig}) may primarily be attributed to different levels of magnetic activity of the sample stars.
Additionally, several other factors could contribute to the dispersion.
For example, despite the cleaning process, the sample might still contain contamination from non-dwarf objects, such as young stellar objects or binaries.
Uncertainties in UV flux measurements could also play a role. 
UV Flares with various rates and strengths \citep{2023ApJ...955...24R} may increase the observed flux with different levels, adding to the scatter.
In some cases, the photospheric emission appears higher than the observed UV emission, likely due to sample contamination, uncertainties in flux measurements, or errors in photospheric flux estimation.

\begin{table}[b]
 \caption{Ratio of photospheric UV emission to observed UV emission for different types of stars. \label{uvph.tab}}
 \begin{center}
 \setlength{\tabcolsep}{3mm}
  \begin{tabular}{cccc}
 \hline\noalign{\smallskip}
 Type & 50\% & 10\% & 90\% \\
  \hline\noalign{\smallskip}
\multicolumn{4}{c}{NUV/GALEX} \\
\noalign{\smallskip}\hline
F & 5.23$\times10^{-1}$ & 3.03$\times10^{-1}$ & 8.51$\times10^{-1}$\\
G & 5.06$\times10^{-1}$ & 2.76$\times10^{-1}$ & 8.35$\times10^{-1}$\\
K & 3.35$\times10^{-1}$ & 9.72$\times10^{-2}$ & 6.79$\times10^{-1}$\\
M & 5.59$\times10^{-2}$ & 4.57$\times10^{-3}$ & 2.38$\times10^{-1}$\\
 \hline\noalign{\smallskip}
\multicolumn{4}{c}{FUV/GALEX} \\
\noalign{\smallskip}\hline
F & 2.49$\times10^{-1}$ & 6.37$\times10^{-2}$ & 4.95$\times10^{-1}$\\
G & 2.34$\times10^{-3}$ & 1.83$\times10^{-4}$ & 1.27$\times10^{-1}$\\
K & 6.67$\times10^{-6}$ & 1.25$\times10^{-7}$ & 1.67$\times10^{-4}$\\
M & 1.73$\times10^{-7}$ & 1.09$\times10^{-8}$ & 2.15$\times10^{-6}$\\
 \hline
\end{tabular}
\end{center}
 {NOTE. The columns 50\%, 10\%, and 90\% represent the 50th, 10th, 90th percentiles of the ratio distribution.}
\end{table}

\subsection{Fitting for CSST bands}

CSST, the space-borne optical-UV telescope, is designed with a primary mirror with a diameter of 2 meters \citep{2023RAA....23g5009J}.
It is equipped with seven photometric imaging bands (i.e., NUV, $u$, $g$, $r$, $i$, $z$, and $y$) and three spectroscopic bands (i.e., GU of 2558--4234 \AA, GV of 3807--6670 \AA, and GI of 6047--10096 \AA), covering a wide range of wavelengths from NUV to near-infrared \citep{2011SSPMA..41.1441Z}. 
CSST offers both a large field of view (FOV) of approximately 1.1 deg$^{2}$ and a high spatial resolution of $0.15^{"}$ for photometric imaging.
In the NUV band, CSST covers a wavelength range from 2481 to 3273 \AA, slightly longer (redder) than the GALEX NUV band. However, it has a detection limit of about 25 mag, significantly deeper than that of the GALEX telescope.
As reported in \citet{2024ApJ...966...69L}, CSST observations hold great potential for conducting UV activity studies, particularly for faint stars that fall below the detection limit of previous and current telescopes. 
To facilitate such research, we also performed a fitting of the photospheric emission level for different types of stars in the CSST NUV band (see Figure \ref{fit.fig} and Table \ref{nuvfit.tab} and \ref{nuvfitc.tab}).

\section{Summary}
\label{sum.sec}

In this paper, we aimed to present a straightforward method to estimate stellar photospheric emission in UV bands.
By comparing observations and various models (e.g., BASTI, PARSEC, MIST, BT-SETTL), we found PARSEC models are better suited for estimating photospheric emission for F to M stars.
We established relations between NUV/GALEX, FUV/GALEX, and NUV/CSST magnitudes
and effective temperatures and colors for different metallicities using tenth-order polynomials.
By using the fitted photospheric magnitudes, the excess UV flux emitted from the chromosphere can be derived to estimate stellar UV activity.

Using the fitting results, we examined the photospheric contribution to observed UV emission in both NUV/GALEX and FUV/GALEX bands.
For most M stars, the photosphere accounts for less than 20\% of the total NUV emission, suggesting the chromosphere predominantly contributes to the NUV emission. Simultaneously, nearly all M stars have negligible photospheric FUV contribution, indicating almost all their FUV emission originates from the chromosphere.
For K stars,  the photospheric contribution to NUV emission typically ranges from 10\% to 70\%, while the photospheric FUV contribution ranges from $10^{-7}$ to $10^{-4}$ (10th to 90th percentiles).
G stars have unignorable photospheric NUV contribution, ranging from 30\% to 85\%, but low photospheric FUV contribution, between $10^{-4}$ and 10\%.
Finally, for F stars, the photospheric contribution is significant in both NUV and FUV bands, with contributions of 30\%--85\% and 6\%--50\%, respectively.

\section{Acknowledgment}

We thank the anonymous referee for helpful comments and suggestions that have significantly improved the paper. 
We thank Drs. Yang Huang and Haibo Yuan for help discussions.
This work was supported by National Key Research and Development Program of China (NKRDPC) under grant Nos. 2019YFA0405000, Science Research Grants from the China Manned Space Project with No. CMS-CSST-2021-A08, Strategic Priority Program of the Chinese Academy of Sciences undergrant No. XDB4100000, and National Natural Science Foundation of China (NSFC) under grant Nos. 11988101/11933004/12273057/11833002/12090042. 


\clearpage
\appendix
\renewcommand*\thetable{\Alph{section}\arabic{table}}
\renewcommand*\thefigure{\Alph{section}\arabic{figure}}

\section{Polynomial fitting results}

Table \ref{nuvfit.tab} and \ref{nuvfitc.tab} list the polynomial fitting results of the relations between photospheric UV magnitudes, including NUV/GALEX, FUV/GALEX, and NUV/CSST bands, and $T_{\rm eff}$ and BP$-$RP for different metallicities, respectively.

\setcounter{table}{0}

\begin{table*}[b]
 \caption{Polynomial fitting results of the relations between NUV/GALEX, FUV/GALEX, and NUV/CSST magnitudes and $T_{\rm eff}$ for different metallicities. \label{nuvfit.tab}}
 \begin{center}
 \setlength{\tabcolsep}{3mm}
\renewcommand{\arraystretch}{0.85}
  \begin{tabular}{crrrrrr}
 \hline\noalign{\smallskip}
\multirow{4}{*}{Index}& \multicolumn{4}{c}{[Fe/H]} \\
\noalign{\smallskip}\cline{2-5}
 & 0.5 & 0.0 & -0.5 & -1.0 \\     
\cline{2-5}\noalign{\smallskip}
& \multicolumn{4}{c}{$T_{\rm eff}$ fitting range (K)} \\
\noalign{\smallskip}\cline{2-5}\noalign{\smallskip}
 & [2600, 7800] & [2600, 7800] & [2600, 7800] & [2800, 7800] \\     
\noalign{\smallskip}\hline\noalign{\smallskip}
\multicolumn{5}{c}{NUV/GALEX} \\
\noalign{\smallskip}\hline
$a_{0}$ & 0.001036119528 & 0.000187162998 & 0.000975317063 & 0.001713663552\\
$a_{1}$ & -0.051712956423 & -0.004945363899 & -0.050088760351 & -0.091778492981\\
$a_{2}$ & 1.132628747192 & -0.011774523259 & 1.131496990617 & 2.182340303252\\
$a_{3}$ & -14.286707177873 & 2.086669682934 & -14.761228488495 & -30.313518645333\\
$a_{4}$ & 114.446818672848 & -37.177961001398 & 122.691560350978 & 272.142347503465\\
$a_{5}$ & -605.143729776001 & 344.181947958909 & -675.578218992532 & -1648.401146528654\\
$a_{6}$ & 2124.39985173958 & -1944.378742862883 & 2479.450178813863 & 6815.882967355477\\
$a_{7}$ & -4846.649885974455 & 6938.531762811192 & -5934.511103773231 & -18980.43573213608\\
$a_{8}$ & 6800.86787731065 & -15272.155811371527 & 8745.973966974187 & 34046.30979026368\\
$a_{9}$ & -5236.769714055182 & 18894.03357910456 & -7018.828224701861 & -35525.08259564591\\
$a_{10}$ & 1699.787322439972 & -9985.337229177087 & 2277.332145500412 & 16424.677666793825\\
\hline\noalign{\smallskip}
\multicolumn{5}{c}{FUV/GALEX} \\
\noalign{\smallskip}\hline
$a_{0}$ & 0.001497586015 & 0.002732628882 & 0.006485610605 & 0.00320807399\\
$a_{1}$ & -0.070071948566 & -0.129102501036 & -0.338156140804 & -0.169768097598\\
$a_{2}$ & 1.388257698153 & 2.623035210094 & 7.797286576862 & 3.979562956413\\
$a_{3}$ & -14.868862200938 & -29.685754767661 & -104.550207767028 & -54.335524542607\\
$a_{4}$ & 88.566727767287 & 200.744257175332 & 901.267732576409 & 477.727295004723\\
$a_{5}$ & -230.930704079958 & -784.594028063851 & -5209.660217744219 & -2820.751167969967\\
$a_{6}$ & -449.350438011906 & 1318.479276092365 & 20409.09799883059 & 11303.251204501981\\
$a_{7}$ & 5547.161715121014 & 2094.342019000447 & -53394.31537513726 & -30282.814865063512\\
$a_{8}$ & -18293.77970569036 & -14804.048360702955 & 89093.97144509312 & 51790.23154381216\\
$a_{9}$ & 28316.3838034019 & 27604.618512926252 & -85473.44022765073 & -50959.92256833935\\
$a_{10}$ & -17384.64995284246 & -18468.811147405188 & 35819.95616358412 & 21940.6939147776\\
\hline\noalign{\smallskip}
\multicolumn{5}{c}{NUV/CSST} \\
\noalign{\smallskip}\hline
$a_{0}$ & 0.000193650809 & -0.000327735355 & 0.000312287855 & 0.001525449726\\
$a_{1}$ & -0.007449988816 & 0.020853639342 & -0.015697046328 & -0.082336535369\\
$a_{2}$ & 0.102141743815 & -0.580986981079 & 0.344010504404 & 1.976597131851\\
$a_{3}$ & -0.297892182296 & 9.351483256982 & -4.301102540603 & -27.7762093994\\
$a_{4}$ & -8.099213220176 & -96.406420374011 & 33.647180042373 & 252.876418714007\\
$a_{5}$ & 118.456494634457 & 665.429045877646 & -169.352002758147 & -1557.465700328339\\
$a_{6}$ & -792.106696888729 & -3114.045288365727 & 538.840195566887 & 6567.529461283991\\
$a_{7}$ & 3080.94004636459 & 9751.199712498441 & -997.010050188479 & -18709.526557598612\\
$a_{8}$ & -7126.751084673026 & -19535.84975698259 & 791.208929848055 & 34439.342228492955\\
$a_{9}$ & 9082.195622441419 & 22577.67945860154 & 292.340019977973 & -36979.50065219379\\
$a_{10}$ & -4875.218129671646 & -11384.404462882925 & -639.37912366556 & 17620.118603477888\\
\noalign{\smallskip}\hline
\end{tabular}
\end{center}
 {NOTE. This table is available on the website of https://github.com/AstroSong/UVphotosphere for a simple usage.}
\end{table*}

\begin{table*}
 \caption{Polynomial fitting results of the relations between NUV/GALEX, FUV/GALEX, and NUV/CSST magnitudes and color BP$-$RP for different metallicities. \label{nuvfitc.tab}}
 \begin{center}
 \setlength{\tabcolsep}{3mm}
\renewcommand{\arraystretch}{0.85}
  \begin{tabular}{crrrrrr}
 \hline\noalign{\smallskip}
\multirow{4}{*}{Index}& \multicolumn{4}{c}{[Fe/H]} \\
\noalign{\smallskip}\cline{2-5}
 & 0.5 & 0.0 & -0.5 & -1.0 \\     
\cline{2-5}\noalign{\smallskip}
& \multicolumn{4}{c}{BP$-$RP fitting range (mag)} \\
\noalign{\smallskip}\cline{2-5}\noalign{\smallskip}
& [0.30, 4.20] & [0.30, 4.30] & [0.30, 4.50] & [0.30, 3.90] \\     
\noalign{\smallskip}\hline\noalign{\smallskip}
\multicolumn{5}{c}{NUV/GALEX} \\
\noalign{\smallskip}\hline
$a_{0}$ & 0.0 & 0.0 & 0.0 & 0.0\\
$a_{1}$ & 0.0 & 0.0 & 0.0 & 0.0\\
$a_{2}$ & -0.096624248688 & 0.0 & 0.012954384272 & -0.007794749635\\
$a_{3}$ & 1.783585096745 & 0.0 & -0.312129002049 & 0.025813714905\\
$a_{4}$ & -13.508985096144 & 0.169855100878 & 3.100531087831 & 0.715464676875\\
$a_{5}$ & 53.757113280351 & -2.480614658371 & -16.495931097307 & -6.866235847797\\
$a_{6}$ & -119.320367997434 & 14.013956376083 & 51.003414794333 & 26.806865906373\\
$a_{7}$ & 144.779035049451 & -38.012567394914 & -92.558030049807 & -55.333398730714\\
$a_{8}$ & -89.410342333598 & 48.841894902942 & 92.744747490896 & 61.282399404072\\
$a_{9}$ & 34.495113922668 & -16.892936078372 & -34.602419547226 & -22.446082155341\\
$a_{10}$ & -0.226864785615 & 6.856181447544 & 9.619905289945 & 8.100040838968\\
\hline\noalign{\smallskip}
\multicolumn{5}{c}{FUV/GALEX} \\
\noalign{\smallskip}\hline
$a_{0}$ & 0.09091119453 & 0.0 & 0.0 & 0.0\\
$a_{1}$ & -2.144199237031 & 0.0 & 0.0 & 0.0\\
$a_{2}$ & 21.640860673039 & -0.180048099617 & 0.013965342826 & -0.173216172433\\
$a_{3}$ & -121.839964230105 & 3.311761557567 & -0.451754899621 & 2.519252604438\\
$a_{4}$ & 418.510930312007 & -24.767606289289 & 5.413558604409 & -14.050981615492\\
$a_{5}$ & -903.043835557777 & 96.315150681636 & -32.690515686016 & 35.641628998721\\
$a_{6}$ & 1220.247843071011 & -206.724207427754 & 109.541709845021 & -30.064448680625\\
$a_{7}$ & -1005.58545485384 & 240.546128545737 & -206.082702349549 & -38.339076260001\\
$a_{8}$ & 477.34389117459 & -142.650829223428 & 204.788301570786 & 92.600662162489\\
$a_{9}$ & -99.311462629693 & 55.735262833695 & -77.178023087759 & -40.642581007802\\
$a_{10}$ & 15.399600526876 & 0.153195716981 & 18.550004180784 & 14.079517848193\\
\hline\noalign{\smallskip}
\multicolumn{5}{c}{NUV/CSST} \\
\noalign{\smallskip}\hline
$a_{0}$ & 0.0 & 0.0 & 0.0 & 0.0\\
$a_{1}$ & 0.0 & 0.0 & 0.0 & 0.0\\
$a_{2}$ & -0.081240936177 & 0.0 & 0.020970246034 & 0.064811578491\\
$a_{3}$ & 1.504639287969 & 0.0 & -0.431009225343 & -1.156114845691\\
$a_{4}$ & -11.424502819072 & 0.100538800206 & 3.684510428498 & 8.546838063748\\
$a_{5}$ & 45.569882972352 & -1.459374102797 & -16.976696190346 & -33.861290965966\\
$a_{6}$ & -101.446365452036 & 8.269963526182 & 45.777270532192 & 77.940187039943\\
$a_{7}$ & 123.311495293858 & -22.871543516603 & -73.538909998519 & -106.535055711987\\
$a_{8}$ & -75.133033827712 & 30.615088613289 & 67.428296303356 & 84.680483656504\\
$a_{9}$ & 26.590913109725 & -10.635975751773 & -23.726443111275 & -28.006119295815\\
$a_{10}$ & 0.28807232791 & 5.417345264004 & 7.417424556965 & 8.044254970785\\
\noalign{\smallskip}\hline
\end{tabular}
\end{center}
 {NOTE. This table is available on the website of https://github.com/AstroSong/UVphotosphere for a simple usage.}
\end{table*}

\end{document}